\documentclass{article}
\usepackage[utf8]{inputenc}

\PassOptionsToPackage{numbers, compress}{natbib}
\usepackage[T1]{fontenc}    \usepackage[british]{babel} \usepackage{csquotes}       \usepackage{hyperref}       \usepackage{booktabs}       \usepackage{amsmath}        \usepackage{amssymb}        \usepackage{amsthm}         \usepackage{amsfonts}       \usepackage{dsfont}         \usepackage{graphicx}       \usepackage{mathtools}      \usepackage{microtype}      \usepackage{mleftright}     \usepackage{multirow}       \usepackage{nicefrac}       \usepackage{paralist}       \usepackage{subcaption}     \usepackage{siunitx}        \usepackage{url}            \usepackage{xspace}         \usepackage{xcolor}

\newcommand{\manifold}{\mathcal{M}}

\newcommand{\reals}   {\mathds{R}}

\DeclareMathOperator{\vietoris}{\mathfrak{V}}

\usepackage{array, makecell}
\graphicspath{ {Figures/} }
\usepackage[inline]{enumitem}
\setlist[enumerate,1]{label=\textit{\alph*)}}
\usepackage{hyperref}

\hypersetup{
    colorlinks=true,
    linkcolor=blue,
    filecolor=magenta,      
    urlcolor=blue,
}

\usepackage{pgfplots}          \usepackage{tikz}              

\usetikzlibrary{calc}

\definecolor{cardinal}{RGB}{196, 30, 58}
\definecolor{bleu}    {RGB}{ 49,140,231}

\pgfplotsset{compat=1.16}

\usepgfplotslibrary{colorbrewer}
\usepgfplotslibrary{fillbetween}
\usepgfplotslibrary{groupplots}

\usepackage[subrefformat=parens]{subcaption}
 \usepackage[final]{neurips_2020}

\usetikzlibrary{external}
\def\sfb{\sffamily \bfseries}

\title{Topological Data Analysis of copy number alterations in cancer}
\author{\textbf{Stefan Groha\textsuperscript{$^{\text{\sfb 1, \sfb 2, \sfb 3,}\dagger}$}} 
  \and \textbf{Caroline Weis\textsuperscript{$^{\text{\sfb 4,\sfb 5,}\dagger}$}}
  \and Alexander Gusev\textsuperscript{$^{\text{\sfb 1, \sfb 2, \sfb 3}}$}
  \and Bastian Rieck\textsuperscript{$^{\text{\sfb 4,\sfb 5}}$} \\
  \small{$^{\text{\sf 1}}$Dana Farber Cancer Institute}; 
  \small{$^{\text{\sf 2}}$Harvard Medical School};
  \small{$^{\text{\sf 3}}$Broad Institute of MIT and Harvard}\\
  \small{$^{\text{\sf 4}}$Machine Learning and Computational Biology Lab, D-BSSE, ETH
        Zurich, Switzerland}\\
  \small{$^{\text{\sf 5}}$SIB Swiss Institute of Bioinformatics, Switzerland}\\
  \small{\texttt{stefanm\_groha@dfci.harvard.edu}; \texttt{caroline.weis@bsse.ethz.ch}}\\
  \small{\textsuperscript{$\dagger$}These authors contributed equally}
 }
\date{September 2020}

\begin{document}

\maketitle
\begin{abstract}
Identifying subgroups and properties of cancer biopsy samples is a crucial step towards obtaining precise diagnoses and being able to perform personalized treatment of cancer patients.
Recent data collections provide a comprehensive characterization of cancer cell data, including genetic data on copy number alterations (CNAs).
We explore the potential to capture information contained in cancer genomic information using a novel topology-based approach that encodes each cancer sample as a persistence diagram of topological features, i.e., high-dimensional voids represented in the data.
We find that this technique has the potential to extract meaningful low-dimensional representations in cancer somatic genetic data and demonstrate the viability of some applications on finding substructures in cancer data as well as comparing similarity of cancer types.
\end{abstract}

\section{Introduction}

Copy number alterations (CNA) are structural somatic mutations where parts of the genome gets repeated or lost. They are common in most cancers~\citep{beroukhim2010landscape} and are known to be involved in cancer development and progression~\citep{budczies2016pan,zhou2017integrated,huang2017copy}. Specific CNAs, as well as the total number of CNAs, were shown to be prognostic for overall survival of cancer patients~\citep{zhang2018association,hieronymus2018tumor,shao2019copy,ried2019landscape}, underlining the importance of these somatic mutations. Furthermore, it has been shown that cancer types can be inferred based on the CNA landscape with simple classification algorithms~\citep{soh2017predicting,zhang2016classification}. Differences in CNAs between lung cancer subtypes, for instance, have also been observed~\citep{li2014classification}, hinting at an inherent structure of these tumor features based on \emph{type} and \emph{subtype} of the underlying tumor. Being able to distinguish and correctly diagnose cancer types and subtypes is of great clinical importance and finding latent biological subgroups or characterizing unknown tumor samples is a question of active research~\citep{bindea2013spatiotemporal,10.1001/jamaoncol.2019.3985,soh2017predicting,sondergaard2017prediction}.

Topological data analysis (TDA) is a tool to study the shape of point clouds, characterizing the underlying physical structure in a representation that is robust to noise, and flexible with respect to selecting a metric. It has gained popularity in recent years due to the fact that it can facilitate the analysis of high-dimensional data by generating meaningful low-dimensional representations in terms of topological features~\citep{Reininghaus15, Hofer17, Moor19Topological}. Methods of computational topology have been applied to CNA comparative genomic hybridization microarray data in breast cancer to distinguish treatment response~\citep{dewoskin2010applications}, identify CNAs that are associated with or predict breast cancer sub-types~\citep{arsuaga2015identification,gonzalez2020prediction}.

We present a proof-of-concept study in which we apply TDA to cancer CNA data to find representations of CNAs per cancer type and characterize similarity between cancer types based on abstracted topological features of their CNA. Furthermore, we examine if the pan-cancer CNA data set factorizes into connected components and if cancer types have meaningful subgroups based on their CNA profile.

\section{Methods}

\subsection{Topological data analysis}

TDA is a recently-emerging field that aims to bring methods from algebraic
and differential topology to machine learning~\citep{Edelsbrunner02}. In contrast to more geometrical techniques, TDA focuses on connectivity
information, which is coarse but also more impervious to noise in the data~\citep{Cohen-Steiner07}. The flagship algorithm of TDA is
called \emph{persistent homology}, a technique for assigning point clouds a set of topological descriptors.
More precisely, with the underlying assumption being that the data $\mathcal{X} :=\{x_1, x_2, \dots, x_n\}$
constitutes a discrete sample from a high-dimensional manifold~$\manifold$, persistent homology analyses
topological features of the point cloud $\mathcal{X}$ at all possible scales. This results in a set of
topological descriptors, the \emph{persistence diagrams}, containing tuples from $\reals \times \reals$ that
describe the ``creation'' and ``destruction'' of topological features.

\begin{figure}
    \centering
    \subcaptionbox{}{\includegraphics[width=0.16\linewidth]{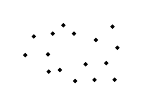}}
    \subcaptionbox{}{\includegraphics[width=0.16\linewidth]{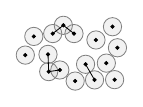}}
    \subcaptionbox{}{\includegraphics[width=0.16\linewidth]{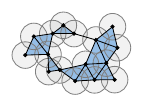}}
    \subcaptionbox{}{\includegraphics[width=0.16\linewidth]{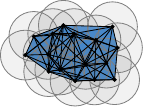}}
    \subcaptionbox{}{\includegraphics[width=0.14\linewidth]{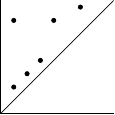}}
    \caption{
        An illustration of the Vietoris--Rips complex construction process. Starting from a set of points
        $\mathcal{X}$, we grow a set of solid spheres (closed metric balls) of radius $\epsilon$ and start connecting points whenever
        two of the solid spheres have a non-empty intersection~(analogously, for subsets satisfying pairwise intersections,
        we create $k$-cliques). While we increase $\epsilon$, we keep track of topological features and mark their
        creation and destruction in a \emph{persistence diagrams}~(right-most plot). The distance of any point to
        the diagonal in this diagram signifies its \emph{persistence}, i.e., its prominence. In this example, we
        observe one high-persistence point, namely the one corresponding to the large-scale cycle or hole. Figure
        is modeled after \citet{Moor19Topological}.
    }
    \label{fig:Overview}
\end{figure}

The advantage of such a description is that the resulting features are interpretable in low dimensions~$d$,
corresponding to \emph{connected components}~($d = 0$) and \emph{cycles}~($d = 1$) in the data, respectively.
Persistent homology supports working with arbitrary similarity measures and metrics, making it a highly
flexible tool for our proposed analysis. We will subsequently calculate the Vietoris--Rips complex $\vietoris(\mathcal{X})$
of the input data~\citep{Vietoris27}, subject to different similarity measures. Figure~\ref{fig:Overview} depicts the calculation;
for illustrative purposes, we used the Euclidean distance here.

\subsection{Copy number alteration (CNA) dataset}
We are basing the analysis on the AACR Project GENIE~(Genomics, Evidence, Neoplasia, Information, Exchange)~\citep{genie} data collection.
This database provides a comprehensive resource for genetic cancer analysis by linking tumor genome information with longitudinal clinical outcomes.
In this work, we base our analysis on copy number alteration~(CNA) values, i.e., somatic changes resulting in multiplication or loss of DNA sections, which are prevalent in many types of cancer.
The copy number alteration values are $\in \{-2, -1.5, -1, 0, 1, 2\}$, leading to a point cloud with discrete values;
it would be formally justified to treat it as a subset of some $\reals^d$, but this structure is also a perfect use case
for employing different similarity measures.
CNA values were determined for a number of genes known to be related to cancer. 
As each hospital determined the CNA values for a slightly different set of genes, and to avoid missing values, we focus on a subset of patients with information present for the same genes, all sequenced at Memorial Sloan Kettering Cancer Center. We furthermore restrict to the largest cancer types with more than 500 samples each.

\section{Experiments}

\subsection{Extraction of topological features per cancer type}
\begin{figure}[tbp]
    \centering
\includegraphics[width=0.9\linewidth]{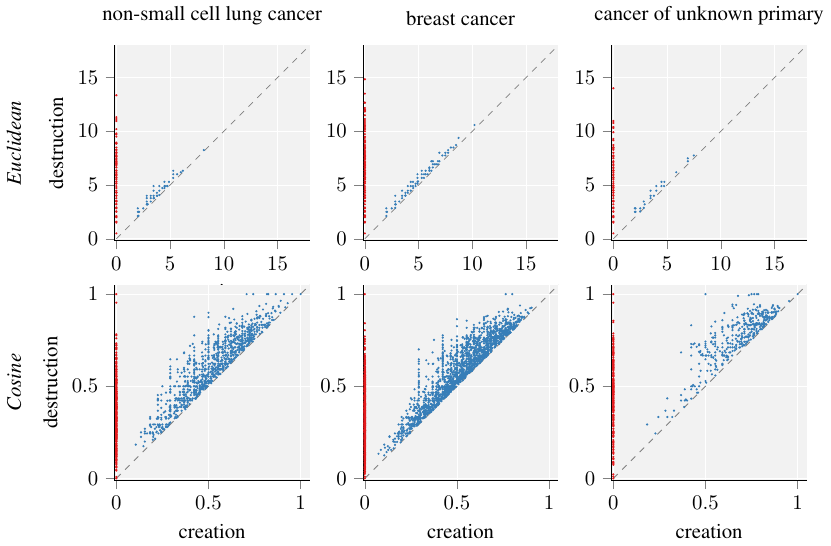}
\caption{
        Persistence diagrams of cancer types non-small cell lung cancer, breast cancer and cancer of unknown primary. Points indicating connected components are depicted in red, point indicating cycles in blue. For each type, the diagram calculated with euclidean distance is depicted in the first row, and cosine distance in the second row. 
              }
    \label{fig:persistence}
\end{figure}

We calculated persistence diagrams on CNA data subsets for each cancer type separately.
We applied two metrics to determine the distance between samples, \emph{Euclidean} and \emph{Cosine} distance.
Formally, the \emph{Cosine} distance just constitutes a similarity measure, lacking some properties required for
it to be a metric, but the persistent homology calculations are still well-defined.
For three cancer types, the resulting persistence diagrams are depicted in Figure~\ref{fig:persistence}.
We observe that a higher number of topological features is detected when using \emph{Cosine} than by using the \emph{Euclidean}
distance.
Their distance to the diagonal indicates that the discovered features are also more persistent than those defined through the \emph{Euclidean} distance.
This reflects the necessity of choosing a domain-specific similarity measure for such an analysis.

We observe gaps between points of dimension~$0$, indicating that well-separated connected components are detected through TDA. 
We hypothesize that these connected components stem from cancer subtypes, separated using a topological lens for the feature space;
a more in-depth investigation of this will require topology-driven clustering algorithms~\citep{Chazal13}.
 A full overview of all persistence diagrams, illustrating the differences over different cancer types, can be found in Appendix Figure~\ref{fig:persistence_suppl}.

\begin{figure}[tb]
    \centering
\includegraphics[width=0.7\linewidth]{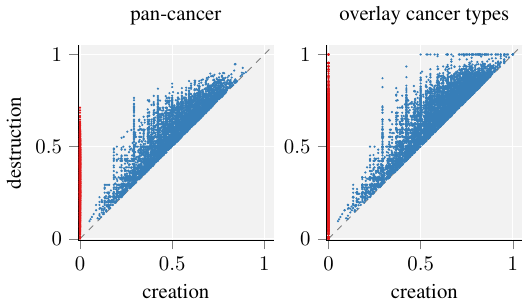}
\caption{
                Comparison of persistence diagrams obtained through extraction of topological features using \emph{Cosine} distance; from the whole dataset, combining all cancer types (left), and overlay of all persistence diagrams calculated on seperate cancer type datasets (right).
              }
    \label{fig:pancancer}
\end{figure}

We furthermore study the independence of all cancer types by combining the persistence diagrams of all cancer types analyzed separately, and comparing this union to the persistence diagram obtained from the full pan-cancer data point-cloud. This is shown in Figure~\ref{fig:pancancer}. We observe more structure when overlaying the separate cancer types,
which points to the fact that the cancer types are somewhat overlapping in the high-dimensional CNA space; data points in overlapping, denser regions are assigned to less persistent features and are therefore not available for larger structures. 

\subsection{Comparison of different cancer types}
\begin{figure}[tbp]
    \centering
\includegraphics[width=0.7\linewidth]{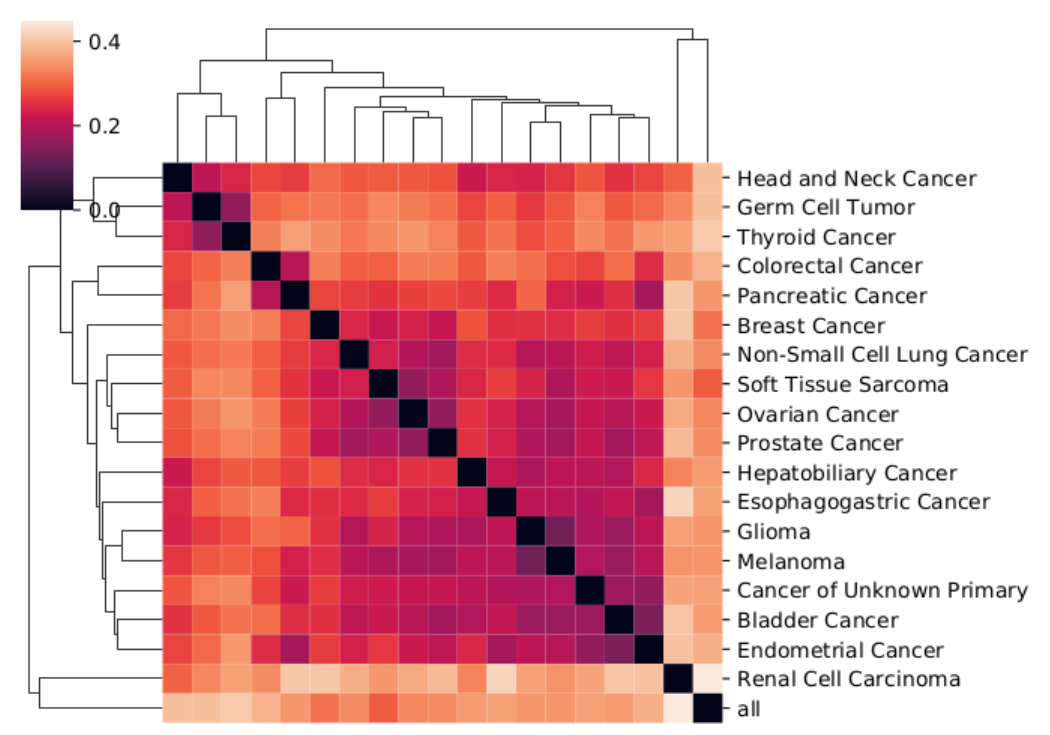}
\caption{
        Heatmap depicting the \emph{bottleneck} distance between persistence diagrams of different cancer types. The persistence diagram calculated from the complete dataset, including all cancer types, is seen in \emph{pan-cancer}.
             }
    \label{fig:heatmap}
\end{figure}

In addition, we compare the persistence diagrams to analyze whether the topological representations are capable of uncovering similarities between different cancer types.
A suitable choice of distance metric between sub-diagrams of persistence diagrams, with the same homological dimension, has to be independent from the different number of detected features.
Different metrics are used to compare pairs of persistence diagrams, based on \emph{heat diffusion}~\citep{Reininghaus15}, \emph{Wasserstein} distance, or the commonly-used \emph{bottleneck} distance; both the Wasserstein distance and the bottleneck distance are instances of metrics
based on optimal transport, with the latter one being more robust towards noise.
Since we observed a clear dependency on the sample size for the \emph{heat diffusion} kernel and the \emph{Wasserstein} distance, we chose to apply the \emph{bottleneck} distance for the analysis.
The resulting distance matrix is depicted in Figure~\ref{fig:heatmap}.
Renal cell carcinoma clearly stands out over other cancer types.
The persistence diagram calculated on the whole dataset, \emph{pan-cancer}, differentiates itself from all single cancer types. 

\section{Conclusion}

We have shown that topological data analysis is a feasible tool to find structure and meaningful representations in tumor copy number alteration data. We have demonstrated that it is possible to  detect subgroups of cancer types based on the connected components of the CNA point cloud, which we hypothesize to be subtypes. Furthermore, we have examined the similarity of cancer types based on topological features extracted from the CNA profile and observe some cancer types to be highly different. This study is a proof-of-concept of the viability of topological data analysis on cancer data and further in-depth analyses are required. Moreover, while only CNA values are used for this study, there are many other somatic mutations (e.g., SNVs, fusions) providing further information on the underlying representation of cancer, possibly better characterizing cancer types and giving the prospect of extending the current analysis to more modalities in future work.

\bibliographystyle{abbrvnat}
\bibliography{final}
  
\newpage
\section*{Appendix}

\begin{figure}[!hb]
    \centering
\includegraphics{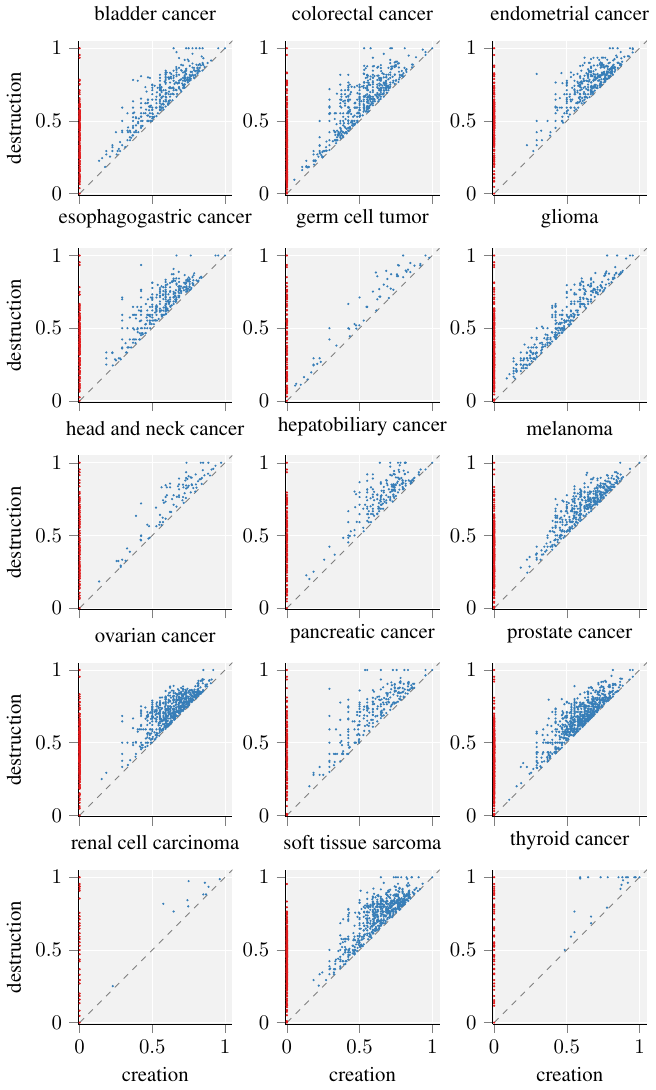}
\caption{
        Persistence diagrams of different cancer types using \emph{Cosine} distance. Points indicating connected components are depicted in red, point indicating cycles in blue. 
              }
    \label{fig:persistence_suppl}
\end{figure} 
\end{document}